# Analytic tomography of the mantle in a spherically Earth. The technique MZY

Yoël Lana-Renault

*Geophysics. Department of Theoretical Physics.*
*University of Zaragoza. 50009 Zaragoza. Spain*
e-mail: *yoel@kepler.unizar.es*

**Abstract.** An explicit expression for P-wave velocity is proposed to develop a novel tomographic technique in a spherically symmetric model of the Earth (MZY). The distribution of the P velocity structure in the mantle is determined using only 34 P- and 2 PcP- observed traveltimes. By applying a non-linear inversion, the P-residuals in the range between 0° and 100° are minimised up to a maximum value of 0.015 s. Furthermore, from the high quality computation of PcP traveltimes, with residuals much better than 0.13 s., it is possible to infer the existence of a brief low velocity layer in the D″ region. This is then followed by a gradual increasing in the velocity profile towards the core, which begins at a depth of 2893.9 km.

**Introduction**

To date, numerous studies use the arrival times of seismic waves to explore the Earth structure. Seismic arrival times have provided a fundamental constraint on the radial and lateral velocity structure of our planet. Sengupta and Toksoz (1976), Clayton and Comer (1983), Dziewonski (1984) among others, studied the variation of the P-wave velocity in the lower mantle. These works have been extended rapidly to the whole mantle (Pulliam et al., 1993) from many different viewpoints and perspectives, but concluding in almost all cases in interesting correlations with the structure predicted by the plate tectonics. On the other hand, reference models constitute the common basis for all the different studies concerning the Earth. Some of them are fairly relevant and well known in the seismological literature, as PREM (Dziewonski and Anderson, 1981), IASP91 (Kennet and Engdahl, 1991) and SP6 (Morelli and Dziewonski, 1993). They constitute the starting point for a number of applications, including seismic tomography and synthetic seismogram calculations. The strategy of finding an agreement between physical meaningful and achieving observations is of crucial importance. A decreasing of the relative error between the reproduced and measured data becomes in an increasing knowledge of the main features concerning the Earth structure. In this sense, any effort made to improve the available reference models, will benefit on the current seismological knowledge, especially those concerning local deviations in boundary interfaces in the Earth's interior.

From this viewpoint, in this preliminary work we pretend to improve the fitting of reference traveltime tables (JB: Jeffreys and Bullen, 1958; BSSA: Herrin et al., 1968) to observed traveltimes and, as consequence of that, to infer the slight deviations of the whole structure with respect to the average models. We have focused our attention on the tomography of the mantle, using and developing a non-linear inversion technique based on the analytical solution of the elliptical integrals involved in the theory of wave propagation. In this sense the approach described in this paper cannot be viewed like an empirical model. We demonstrate that the range of the achievement is large enough and, therefore, the real interpretation is to be an improvement for the reference model derived from the Herrin et al. traveltime tables, used in this work. Eventually, this sort of agreement to respect the *observed* data (errors not larger than a particular threshold) has been imposed as a first objective of this study, but it is not unique. The use of an analytical function avoids the common strategy of deriving spherical averages from seismological observations via an inversion procedure (i.e., the least-square approach). An interesting comment of this performing can be found in Morelli and Dziewonski (1993). In our scheme, the inherent biased data distribution is largely overcome since only traveltime tables are taking into account. This absence of real data is a major lack in the model we present in this paper, and we agree. However, we keep the opinion that the results should be interpreted in a different way as those derived from a reference model, because they maintain internal consistency and do not pretend to be an alternative to PREM, ISAP91 or SP6 models.

The use of analytical functions to derive a model that globally reproduces the observed traveltimes by acting locally on a multilayer and spherical mantle does not prescribe the meaning, from a physical viewpoint, of the new model. Indeed, the analytical tomography results in an improved understanding of some particular areas, for example the $D''$ layer at the base of the mantle. These features are the most relevant conclusions of our work as they provide some slight differences to the current knowledge of the mantle.

**Methodology**

The trial P-wave velocity function used in this work to analyse the structure of the mantle can be summarised by the expression

$$v(r) = r \cdot (B - A \cdot \ln(r)) \quad , \quad (1)$$

where $r$ is the radius, and $(A,B)$ two independent parameters to be determined. This formula can be simplified by defining the function

$$w(r) = (B - A \cdot \ln(r)) \quad , \quad (2)$$

and then:

$$v(r) = r \cdot w(r) \quad . \quad (3)$$

The P-wave velocity function expressed in Eq. (1) has been used (Lana-Renault and Cid, 1991; Lana-Renault, 1998) to obtain different Earth models by varying the different parameters. The smoothness of this function makes it adequate to tomographic studies of the Earth's mantle, once a proper parameterisation is applied, i.e. a division in many spherical layers, which is the one followed in this work. Another useful property of the function described in Eq. (1) is that converts the elliptical integrals arisen during the hamiltonian formulation of ray propagation, into analytical functions. For example, for a ray crossing the first layer ($i = 1$) of the mantle, who radius of the top surface is $R_1$ (Earth's radius), the epicentral distance $D$ can be expressed as

$$D = \frac{2 \cdot w_1}{A_1} \cdot \operatorname{senh}\left(\frac{A_1 \cdot T}{2}\right) \quad , \quad (4)$$

where:

$$w_1 = w_1(R_1) = B_1 - A_1 \cdot \ln(R_1) \quad .$$

The general analytical expressions for $D$ and $T$ can be obtained using the classical integral expressions (Bullen and Bolt, 1985)

$$D = p \cdot \int_{r_p}^{r_o} r^{-1} \cdot \sqrt{h^2 - p^2} \, dr \quad ,$$

$$T = \int_{r_p}^{r_o} h^2 \cdot r^{-1} \cdot \sqrt{h^2 - p^2} \, dr \quad .$$

(5)

Denoting the angle of incidence at the top surface of the $i^{th}$ layer by $I_i$ and its radius by $R_i$, and similarly for the variables at the bottom ($I'_i$ and $R'_i$), see figure 1, it is always possible to write

$$w_i = w_i(R_i) = B_i - A_i \cdot \ln(R_i) = \frac{sen(I_i)}{p} \quad , \quad (6)$$

and

$$w'_i = w'_i(R'_i) = B_i - A_i \cdot \ln(R'_i) = \frac{sen(I'_i)}{p} \quad . \quad (7)$$

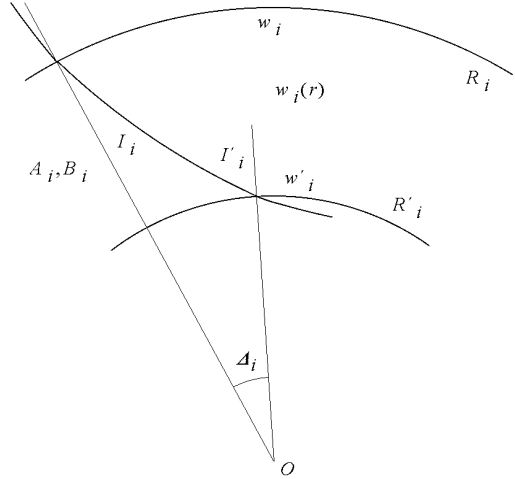

**Fig. 1.** P-trajectory traveling through a layer *i*

In general, for a point *P(r)* we have

$$w_i(r) = B_i - A_i \cdot \ln(r) = \frac{sen(I)}{p} \quad . \quad (8)$$

Hence, through its derivative,

$$\frac{dr}{r \cdot \cos(I)} = -\frac{dI}{p \cdot A_i} \quad (9)$$

we can calculate the expressions (5) for a P-trajectory which travels from $R_i$ to $R'_i$

$$D_i = \frac{\cos(I_i) - \cos(I'_i)}{p \cdot A_i} \quad , \quad (10)$$

$$T_i = \frac{1}{A_i} \cdot \ln \left[ \frac{tg\left(\frac{I'_i}{2}\right)}{tg\left(\frac{I_i}{2}\right)} \right] \quad . \quad (11)$$

Therefore, the observables at the Earth surface can be computed as a result of several additions of these computed values at each layer. That is, if one ray travels along *k* layers, the final epicentral distance and traveltime are calculated through of the following *2k+1* equations:

$$D = \frac{2}{p} \cdot \left( \sum_i \frac{\cos(I_i) - \cos(I´_i)}{A_i} + \frac{\cos(I_k)}{A_k} \right), \quad (12)$$

$$T = 2 \cdot \left[ \sum_i \frac{1}{A_i} \cdot \ln\left[ \frac{tg\left(\frac{I´_i}{2}\right)}{tg\left(\frac{I_i}{2}\right)} \right] - \frac{1}{A_k} \cdot \ln\left( tg\left(\frac{I_k}{2}\right) \right) \right] \quad (13)$$

and these *2k-1* auxiliary equations

$$p = \frac{sen(I_i)}{w_i(R_i)} = \frac{sen(I´_i)}{w_i(R´_i)} = \frac{sen(I_k)}{w_k(R_k)}, \quad (14)$$

where: *i* = 1, 2, ..., *k-1* .

On the other hand, the observables for a PcP-trajectory are calculated by the following *2(k+1)* equations:

$$D = \frac{2}{p} \cdot \left( \sum_i \frac{\cos(I_i) - \cos(I´_i)}{A_i} \right), \quad (15)$$

$$T = 2 \cdot \left[ \sum_i \frac{1}{A_i} \cdot \ln\left[ \frac{tg\left(\frac{I´_i}{2}\right)}{tg\left(\frac{I_i}{2}\right)} \right] \right] \quad (16)$$

and these *2k* auxiliary equations

$$p = \frac{sen(I_i)}{w_i(R_i)} = \frac{sen(I´_i)}{w_i(R´_i)}. \quad (17)$$

where: *i* = 1, 2, ..., *k* .

Finally, by integrating Eq. (9) between $P(R_i)$ and $P(r)$, it is easy to calculate the radius of any single point $P(r)$ along the trajectory:

$$r = R_i \cdot exp\left( \frac{sen(I_i) - sen(I)}{p \cdot A_i} \right) = R_i \cdot exp\left( \frac{w_i - w_i(r)}{A_i} \right). \quad (18)$$

**Results**

With a single collection of observed traveltimes, it is possible to reproduce the observations on the Earth's surface for any event. For the sake of simplicity, as an example of the versatility and functionality of the proposed methodology, we have selected the datasets reproduced in *Herrin et al.* (1968). The sequence of calculations consists of determining the specific constants $A_i$, $w_i$ and $w´_i$ ($i = 1,...,N$), for each layer, $N$ being the number of layers.

(Note that $w´_i$ is a measure of the thickness of the $i^{th}$ layer and that we don't use $B_i$. The parameter $B_i$ is calculated after using the Eq. (6))

Let suppose these quantities are already known for the first $k-1$ layers, except $w´_{k-1}$, the starting point for the $k^{th}$ layer. The inverse problem can be posed as a system of non-linear equations (12-14) that will provide the parameters $A_k$ and $w_k$ of the layer $k$. We must use three P-observed trajectories reproducing three fixed points ($D_l$, $T_{ol}$ ; $l = 1,2,3$) as boundary conditions for the system of $3(2k+1)$ non linear equations with $3(2k+1)$ unknowns ($p_l$, $I_{il}$, $I´_{il}$, $I_{kl}$, $w´_{k-1}$, $w_k$, $A_k$). The solution is then iterated till assure a convergence criterion, in our case, a threshold for the computed residuals less than a certain value ($10^{-15}$).

Known the values $w´_{k-1}$, $w_k$, y $A_k$, we prove that the residual times $T_o$-$T_c$ (observed minus computed time) of the all the others P-trajectories which also return to the surface-focus from the $k^{th}$ layer are smaller than a determined $e$. If it is not so, we begin again taking others three observables ($D_l$, $T_{ol}$) nearer among them.

It is to be noted that the algebra applied in our methodology permits a discontinuity of the 1$^{st}$ kind ($w´_{k-1} \neq w_k$) in the velocity function.

This last property can be analysed through the study of the derivative ($dT/dD$) (*Herrin et al.*, 1968), in order to detect jumps in the selected velocity pattern. If we have the security that only a discontinuity of the 2$^{nd}$ kind ($w´_{k-1} = w_k$) is present, then it is possible to work with only 2 observables (**D**, $T_o$) or fixed boundary conditions, thus eliminating $2k+1$ redundant equations from the global system.

In this case, the experience tells us that is much better to work with one observable (**D**, $T_o$) and, thus, fixing the final of the $k-1^{th}$ layer by a value for $w´_{k-1}$. and insuring that the residual times of all the P-trajectories which return from the $k-1^{th}$ layer to the surface-focus are less than our $e$. Thereby, we resolve a non linear system with only $2k+1$ equations.

We have performed a complete description of the Mantle using a maximum residual time $\varepsilon = 0.015$ s. and only 34 P-observed traveltimes. The total number of layers used in this description is 28. The last one finishes at a depth of 2810.1 km., maximum for the last P-observed trajectory at **D** = 100° according to *Herrin et al.* (1968), with $T_o = 826.7303$ s. Once known the problematic of the lack of information for **D** > 100° and the special case of the D" shell, we have worked with data available from $D > 88°$ and maximum residuals of *0.002 s.* See Table 1 and Residuals of P-travel times (Figure 2) for a graphic representation and further details.

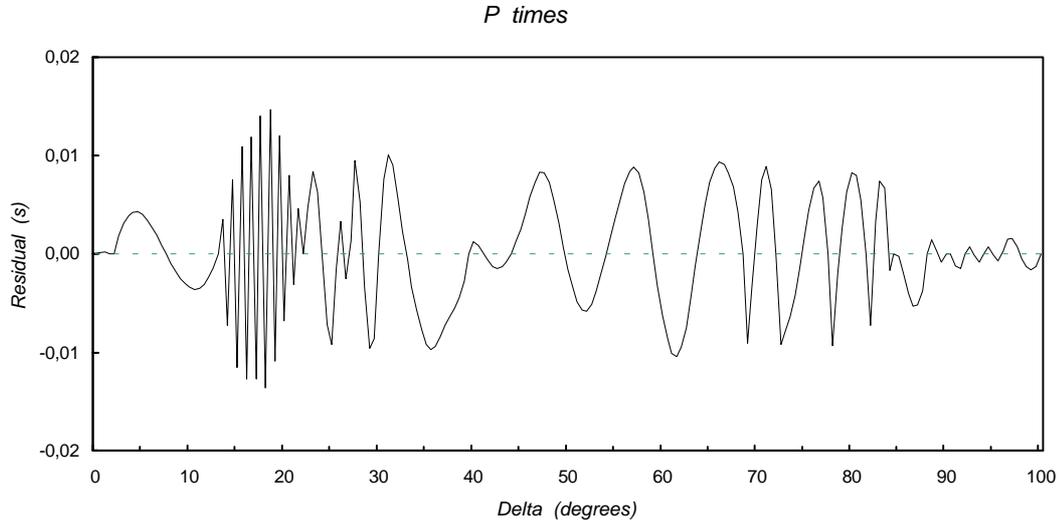

**Fig. 2.** P-residual times $T_o-T_c$. $D: 0 – 100°$ every *0.5°*. Maximum residual *0.015 sec.* at $D = 18.5°$.

*The 29$^{th}$ layer*. Since the derivative associated with the surface-focus travel time (*dT/dD*) is effectively constant (*Herrin et al.*, 1968) beyond 99.0°, and that our residuals are practically zero for those points, we consider the boundary condition $w´_{28} = 1/p(100°) = w_{29}$ produces the best results. Thus, our problem is reduced to calculate the parameters $w´_{29}$ (final layer) and $A_{29}$. For this purpose, we pose a non linear system with two PcP observables. One of these fixed points should always be the axial trajectory ($D = 0°$ ; $T_o = 511.3$ sec.) that allows us to use only one equation:

$$T = 2 \cdot \sum_i \frac{1}{A_i} \cdot \ln\left(\frac{w´_i}{w_i}\right) \qquad (19)$$

We have considered that the other observable should be very separated from the first, and thus, selected $D = 93°$ ; $T_o = 795.2$ s as second observable. Once obtained the values $w´_{29}$ and $A_{29}$, all the PcP residuals were balanced with an error less than 0.13 s. (Figure 3, Table 2 of PcP-travel times). By applying Eq. (18) we found the outer core at 2893.9 km. Further details can be seen in Figures 4, 5, 6 and Table 3 of P-wave velocity.

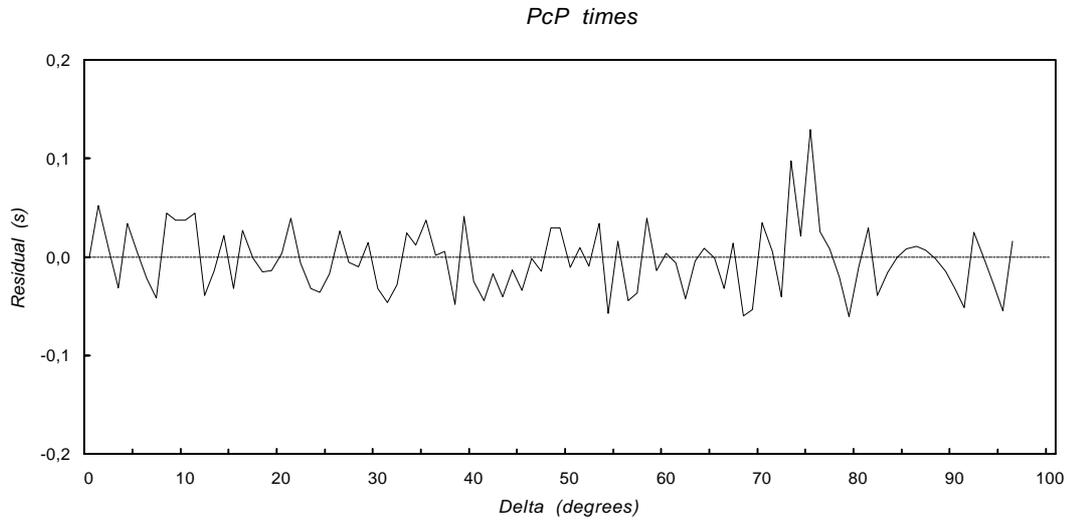

**Fig. 3.** PcP-residual times $T_o$-$T_c$ . ***D**: 0 – 96°* every *1°*. Maximum residual *0.13 sec.* at ***D** = 75°*. Depth Outer Core: *2893.9 km.*

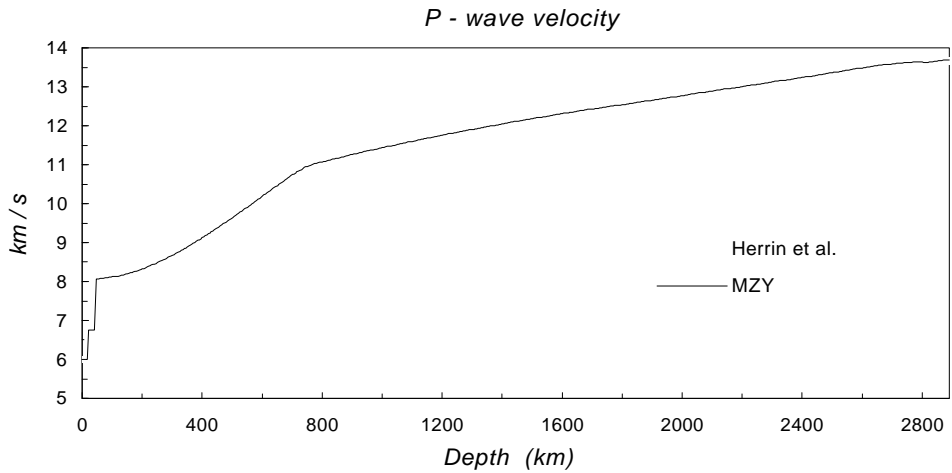

**Fig. 4.** P-wave velocity in the mantle. *Herrin et al.* versus *MZY*.

Fig. 4 shows a comparison between our velocity distribution and the one provided by *Herrin et al.* (1968). The most conspicuous difference is observed at *2749.8 km.*, final of the *25th* layer, where the last trajectory returns to the surface at ***D** = 92°* and the corresponding residual is null. Figure 5 exhibits residuals of computed velocity *Herrin et al.* minus *MZY*. Let us note how the maximum residual is *0.0044 km/s* at a depth of *755 km.*

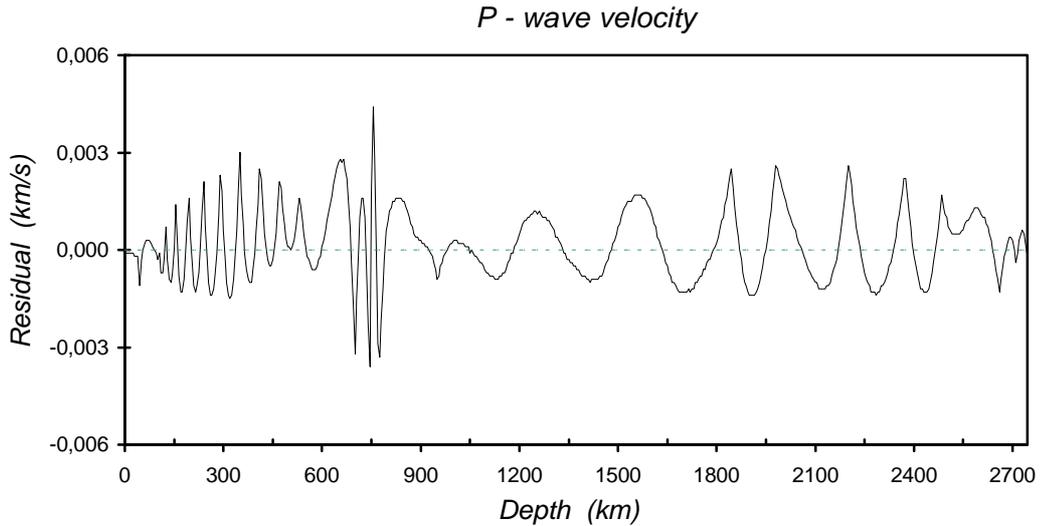

**Fig. 5.** Residual velocity *Herrin et al.* minus *MZY* in the mantle until a depth of *2745 km.* Maximum residual: *0.0044 km/sec* at a depth of *755 km.*

From 2749.8 km. until the core-mantle boundary (*region D"*), our velocity distribution begins to be completely different to *Herrin et al.* (1968), as can be seen in Fig. 6. This is due that *Herrin et al.* adopted a special smooth velocity distribution to the region D" to explain the last results from *Taggart and Engdahl,* (1968), which indicated a slow increase of velocity towards the core. Morelli and Dziewonski (1993), in their SP6 model, obtained a continuous decrease from 2741 km. In our MZY model, we propose that D" region begins at 2780.7 km. with a brief (29.4 km.) negative gradient (layers number 27 and 28) followed by a slow increase until the core. With this profile we insure the residuals of all the observables P and PcP from *Herrin et al.* (1968) are minima, and reproduce accurately the observed times.

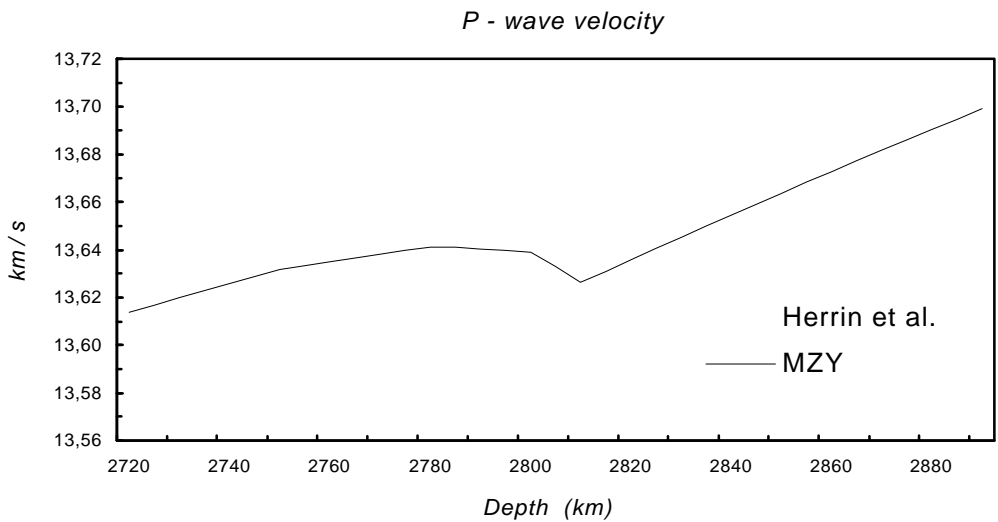

**Fig. 6.** P-wave velocity in D" region. *Herrin et al.* versus *MZY*

## Concluding remarks

We have presented a new technique to tomography the interior of the Earth for which one is able to obtain residual times less than a determined value *e* for all observed trajectories P. The more minor is the value of *e*, more genuine and real is the tomography.

Also, we have seen that the technique MZY developed is very easy to apply. Its potentiality is based in the function velocity found that it provides us analytical solutions for Δ and T.

## Acknowledgements

The author is grateful to Dr. Javier Sabadell by the comments and suggestions made during the writing of the manuscript.

| Table 1 | | | | | P travel times  _(sec.)_ | | | | p. 1/2 |
|---|---|---|---|---|---|---|---|---|---|
| La-yers | $\Delta_{initial}$ $\Delta_{final}$ | $\Delta$ | Observed (Herrin & al. ) | Computed MZY | Residual $T_o - T_c$ | La-yers | $\Delta_{initial}$ $\Delta_{final}$ | $\Delta$ | Observed (Herrin & al. ) | Computed MZY | Residual $T_o - T_c$ |
| 1 | 0 | 0.00 | 0.0000 | 0.0000 | 0.0000 | | 27.14 | 27.50 | 347.2025 | 347.1930 | 0.0095 |
|   |   | 0.50 | 9.2663 | 9.2662 | 0.0001 |   |   | 28.00 | 351.6796 | 351.6743 | 0.0053 |
|   | 7.85 | 1.00 | 18.5323 | 18.5321 | 0.0002 | 15 |   | 28.50 | 356.1456 | 356.1490 | -0.0034 |
| 2 | 0.52 | 1.50 | 26.9525 | 26.9525 | 0.0000 |   |   | 29.00 | 360.6048 | 360.6144 | -0.0096 |
|   | 10.66 |   |   |   |   |   |   | 29.50 | 365.0596 | 365.0682 | -0.0086 |
|   | 0.989 | 2.00 | 34.8630 | 34.8630 | 0.0000 |   | 30 | 30.00 | 369.5086 | 369.5086 | 0.0000 |
|   |   | 2.50 | 41.7231 | 41.7213 | 0.0018 |   | 30 | 30.50 | 373.9477 | 373.9401 | 0.0076 |
|   |   | 3.00 | 48.5813 | 48.5782 | 0.0031 |   |   | 31.00 | 378.3751 | 378.3650 | 0.0101 |
|   |   | 3.50 | 55.4373 | 55.4334 | 0.0039 |   |   | 31.50 | 382.7900 | 382.7810 | 0.0090 |
|   |   | 4.00 | 62.2906 | 62.2864 | 0.0042 |   |   | 32.00 | 387.1923 | 387.1867 | 0.0056 |
|   |   | 4.50 | 69.1410 | 69.1367 | 0.0043 |   |   | 32.50 | 391.5831 | 391.5808 | 0.0023 |
|   |   | 5.00 | 75.9880 | 75.9840 | 0.0040 |   |   | 33.00 | 395.9621 | 395.9627 | -0.0006 |
|   |   | 5.50 | 82.8312 | 82.8278 | 0.0034 |   |   | 33.50 | 400.3281 | 400.3315 | -0.0034 |
|   |   | 6.00 | 89.6703 | 89.6676 | 0.0027 |   |   | 34.00 | 404.6807 | 404.6864 | -0.0057 |
|   |   | 6.50 | 96.5049 | 96.5030 | 0.0019 | 16 |   | 34.50 | 409.0193 | 409.0270 | -0.0077 |
|   |   | 7.00 | 103.3346 | 103.3337 | 0.0009 |   |   | 35.00 | 413.3435 | 413.3527 | -0.0092 |
| 3 |   | 7.50 | 110.1591 | 110.1591 | 0.0000 |   |   | 35.50 | 417.6532 | 417.6629 | -0.0097 |
|   |   | 8.00 | 116.9779 | 116.9789 | -0.0010 |   |   | 36.00 | 421.9479 | 421.9573 | -0.0094 |
|   |   | 8.50 | 123.7908 | 123.7926 | -0.0018 |   |   | 36.50 | 426.2269 | 426.2353 | -0.0084 |
|   |   | 9.00 | 130.5973 | 130.5998 | -0.0025 |   |   | 37.00 | 430.4894 | 430.4967 | -0.0073 |
|   |   | 9.50 | 137.3970 | 137.4000 | -0.0030 |   |   | 37.50 | 434.7347 | 434.7411 | -0.0064 |
|   |   | 10.00 | 144.1896 | 144.1930 | -0.0034 |   |   | 38.00 | 438.9626 | 438.9681 | -0.0055 |
|   |   | 10.50 | 150.9747 | 150.9783 | -0.0036 |   |   | 38.50 | 443.1730 | 443.1775 | -0.0045 |
|   |   | 11.00 | 157.7519 | 157.7554 | -0.0035 |   |   | 39.00 | 447.3662 | 447.3690 | -0.0028 |
|   |   | 11.50 | 164.5209 | 164.5240 | -0.0031 |   | 39.5 | 39.50 | 451.5425 | 451.5425 | 0.0000 |
|   |   | 12.00 | 171.2813 | 171.2836 | -0.0023 |   | 39.5 | 40.00 | 455.7020 | 455.7008 | 0.0012 |
|   |   | 12.50 | 178.0326 | 178.0340 | -0.0014 |   |   | 40.50 | 459.8449 | 459.8440 | 0.0009 |
|   | 13.15 | 13.00 | 184.7746 | 184.7746 | 0.0000 |   |   | 41.00 | 463.9710 | 463.9709 | 0.0001 |
| 4 | 13.15 | 13.50 | 191.4964 | 191.4929 | 0.0035 |   |   | 41.50 | 468.0802 | 468.0808 | -0.0006 |
|   | 14.14 | 14.14 | 200.0582 * | 200.0655 | -0.0073 |   |   | 42.00 | 472.1723 | 472.1736 | -0.0013 |
| 5 | 14.14 | 14.50 | 204.8555 | 204.8480 | 0.0075 |   |   | 42.50 | 476.2473 | 476.2488 | -0.0015 |
|   | 15.13 | 15.13 | 213.1831 * | 213.1946 | -0.0115 |   |   | 43.00 | 480.3051 | 480.3064 | -0.0013 |
| 6 | 15.13 | 15.50 | 218.0429 | 218.0320 | 0.0109 |   |   | 43.50 | 484.3454 | 484.3462 | -0.0008 |
|   | 16.11 | 16.11 | 225.9644 * | 225.9771 | -0.0127 |   |   | 44.00 | 488.3680 | 488.3680 | 0.0000 |
| 7 | 16.11 | 16.50 | 230.9845 | 230.9726 | 0.0119 |   |   | 44.50 | 492.3728 | 492.3716 | 0.0012 |
|   | 17.09 | 17.09 | 238.4697 * | 238.4824 | -0.0127 |   |   | 45.00 | 496.3596 | 496.3571 | 0.0025 |
| 8 | 17.09 | 17.50 | 243.6096 | 243.5956 | 0.0140 |   |   | 45.50 | 500.3285 | 500.3244 | 0.0041 |
|   | 18.08 | 18.08 | 250.7491 * | 250.7627 | -0.0136 |   |   | 46.00 | 504.2791 | 504.2733 | 0.0058 |
| 9 | 18.08 | 18.50 | 255.8408 | 255.8261 | 0.0147 |   |   | 46.50 | 508.2111 | 508.2038 | 0.0073 |
|   | 19.06 | 19.06 | 262.4864 * | 262.4973 | -0.0109 |   |   | 47.00 | 512.1242 | 512.1159 | 0.0083 |
| 10 | 19.06 | 19.50 | 267.6136 | 267.6016 | 0.0120 | 17 |   | 47.50 | 516.0178 | 516.0096 | 0.0082 |
|   | 20.04 | 20.04 | 273.7653 * | 273.7721 | -0.0068 |   |   | 48.00 | 519.8920 | 519.8847 | 0.0073 |
| 11 | 20.04 | 20.50 | 278.9036 | 278.8956 | 0.0080 |   |   | 48.50 | 523.7469 | 523.7415 | 0.0054 |
|   | 21.02 | 21.02 | 284.5832 * | 284.5863 | -0.0031 |   |   | 49.00 | 527.5828 | 527.5797 | 0.0031 |
| 12 | 21.02 | 21.50 | 289.7160 | 289.7114 | 0.0046 |   |   | 49.50 | 531.4001 | 531.3995 | 0.0006 |
|   | 22 | 22.00 | 294.9501 | 294.9501 | 0.0000 |   |   | 50.00 | 535.1992 | 535.2008 | -0.0016 |
|   | 22 | 22.50 | 300.0806 | 300.0759 | 0.0047 |   |   | 50.50 | 538.9802 | 538.9837 | -0.0035 |
|   |   | 23.00 | 305.1134 | 305.1050 | 0.0084 |   |   | 51.00 | 542.7433 | 542.7482 | -0.0049 |
| 13 |   | 23.50 | 310.0533 | 310.0470 | 0.0063 |   |   | 51.50 | 546.4887 | 546.4944 | -0.0057 |
|   |   | 24.00 | 314.9070 | 314.9070 | 0.0000 |   |   | 52.00 | 550.2164 | 550.2222 | -0.0058 |
|   |   | 24.50 | 319.6818 | 319.6890 | -0.0072 |   |   | 52.50 | 553.9266 | 553.9317 | -0.0051 |
|   | 25.46 | 25.00 | 324.3869 | 324.3961 | -0.0092 |   |   | 53.00 | 557.6192 | 557.6230 | -0.0038 |
|   | 25.46 | 25.46 | 328.6614 * | 328.6630 | -0.0016 |   |   | 53.50 | 561.2941 | 561.2962 | -0.0021 |
| 14 |   | 26.00 | 333.6295 | 333.6262 | 0.0033 |   |   | 54.00 | 564.9510 | 564.9512 | -0.0002 |
|   |   | 26.50 | 338.1848 | 338.1873 | -0.0025 |   |   | 54.50 | 568.5899 | 568.5882 | 0.0017 |
|   | 27.14 | 27.14 | 343.9656 * | 343.9643 | 0.0013 |   |   | 55.00 | 572.2107 | 572.2072 | 0.0035 |

| Table 1 | | | | | P travel times (sec.) | | | | | p. 2/2 |
|---|---|---|---|---|---|---|---|---|---|---|
| La-yers | $\Delta_{initial}$ $\Delta_{final}$ | $\Delta$ | Observed (Herrin & al. ) | Computed MZY | Residual $T_o - T_c$ | La-yers | $\Delta_{initial}$ $\Delta_{final}$ | $\Delta$ | Observed (Herrin & al. ) | Computed MZY | Residual $T_o - T_c$ |
| 17 | | 55.50 | 575.8137 | 575.8082 | 0.0055 | | 81.8 | 81.80 | 740.2336 * | 740.2409 | -0.0073 |
| | | 56.00 | 579.3986 | 579.3915 | 0.0071 | | | 82.50 | 743.9007 | 743.8974 | 0.0033 |
| | | 56.50 | 582.9653 | 582.9569 | 0.0084 | 22 | | 83.00 | 746.4926 | 746.4852 | 0.0074 |
| | | 57.00 | 586.5135 | 586.5047 | 0.0088 | | | 83.50 | 749.0611 | 749.0544 | 0.0067 |
| | | 57.50 | 590.0430 | 590.0348 | 0.0082 | | 84.04 | 84.04 | 751.8075 * | 751.8092 | -0.0017 |
| | | 58.00 | 593.5538 | 593.5475 | 0.0063 | | 84.04 | 84.50 | 754.1271 | 754.1271 | 0.0000 |
| | | 58.50 | 597.0462 | 597.0427 | 0.0035 | | | 85.00 | 756.6260 | 756.6263 | -0.0003 |
| | 59 | 59.00 | 600.5205 | 600.5205 | 0.0000 | | | 85.50 | 759.1042 | 759.1061 | -0.0019 |
| | 59 | 59.50 | 603.9770 | 603.9803 | -0.0033 | 23 | | 86.00 | 761.5636 | 761.5676 | -0.0040 |
| | | 60.00 | 607.4162 | 607.4225 | -0.0063 | | | 86.50 | 764.0064 | 764.0117 | -0.0053 |
| | | 60.50 | 610.8385 | 610.8471 | -0.0086 | | | 87.00 | 766.4338 | 766.4390 | -0.0052 |
| | | 61.00 | 614.2444 | 614.2545 | -0.0101 | | | 87.50 | 768.8465 | 768.8502 | -0.0037 |
| | | 61.50 | 617.6343 | 617.6447 | -0.0104 | | 88 | 88.00 | 771.2455 | 771.2455 | 0.0000 |
| | | 62.00 | 621.0084 | 621.0179 | -0.0095 | | 88 | 88.50 | 773.6315 | 773.6301 | 0.0014 |
| | | 62.50 | 624.3668 | 624.3743 | -0.0075 | 24 | | 89.00 | 776.0056 | 776.0053 | 0.0003 |
| | | 63.00 | 627.7094 | 627.7138 | -0.0044 | | | 89.50 | 778.3687 | 778.3695 | -0.0008 |
| | | 63.50 | 631.0356 | 631.0367 | -0.0011 | | 90 | 90.00 | 780.7222 | 780.7222 | 0.0000 |
| 18 | | 64.00 | 634.3452 | 634.3430 | 0.0022 | | 90 | 90.50 | 783.0673 | 783.0673 | 0.0000 |
| | | 64.50 | 637.6379 | 637.6329 | 0.0050 | 25 | | 91.00 | 785.4049 | 785.4061 | -0.0012 |
| | | 65.00 | 640.9137 | 640.9064 | 0.0073 | | | 91.50 | 787.7356 | 787.7371 | -0.0015 |
| | | 65.50 | 644.1724 | 644.1637 | 0.0087 | | 92 | 92.00 | 790.0597 | 790.0597 | 0.0000 |
| | | 66.00 | 647.4142 | 647.4049 | 0.0093 | | 92 | 92.50 | 792.3774 | 792.3767 | 0.0007 |
| | | 66.50 | 650.6392 | 650.6301 | 0.0091 | 26 | | 93.00 | 794.6891 | 794.6892 | -0.0001 |
| | | 67.00 | 653.8477 | 653.8395 | 0.0082 | | | 93.50 | 796.9953 | 796.9961 | -0.0008 |
| | | 67.50 | 657.0398 | 657.0330 | 0.0068 | | 94.03 | 94.03 | 799.4344 * | 799.4344 | 0.0000 |
| | | 68.00 | 660.2151 | 660.2108 | 0.0043 | | 94.03 | 94.50 | 801.5937 | 801.5930 | 0.0007 |
| | | 68.50 | 663.3731 | 663.3731 | 0.0000 | 27 | | 95.00 | 803.8872 | 803.8873 | -0.0001 |
| | 69.08 | 69.08 | 667.0129 * | 667.0220 | -0.0091 | | | 95.50 | 806.1777 | 806.1784 | -0.0007 |
| | 69.08 | 69.50 | 669.6355 | 669.6387 | -0.0032 | | 96.123 | 96.123 | 809.0282 * | 809.0279 | 0.0003 |
| | | 70.00 | 672.7383 | 672.7354 | 0.0029 | | 96.123 | 96.50 | 810.7518 | 810.7503 | 0.0015 |
| 19 | | 70.50 | 675.8202 | 675.8127 | 0.0075 | | | 97.00 | 813.0361 | 813.0345 | 0.0016 |
| | | 71.00 | 678.8805 | 678.8716 | 0.0089 | | | 97.50 | 815.3192 | 815.3185 | 0.0007 |
| | | 71.50 | 681.9193 | 681.9128 | 0.0065 | 28 | | 98.00 | 817.6016 | 817.6021 | -0.0005 |
| | 72.35 | 72.00 | 684.9366 | 684.9366 | 0.0000 | | | 98.50 | 819.8838 | 819.8851 | -0.0013 |
| | 72.35 | 72.35 | 687.0340 * | 687.0432 | -0.0092 | | | 99.00 | 822.1660 | 822.1676 | -0.0016 |
| | | 73.00 | 690.9092 | 690.9170 | -0.0078 | | | 99.50 | 824.4481 | 824.4494 | -0.0013 |
| | | 73.50 | 693.8665 | 693.8729 | -0.0064 | | 100 | 100.00 | 826.7303 | 826.7303 | 0.0000 |
| | | 74.00 | 696.8054 | 696.8096 | -0.0042 | | 100 | 99.5 | 824.4481 | 824.4498 | -0.0017 |
| | | 74.50 | 699.7264 | 699.7279 | -0.0015 | | | 99.0 | 822.1660 | 822.1694 | -0.0034 |
| 20 | | 75.00 | 702.6299 | 702.6283 | 0.0016 | | | 98.5 | 819.8838 | 819.8892 | -0.0054 |
| | | 75.50 | 705.5159 | 705.5115 | 0.0044 | | | 98.0 | 817.6016 | 817.6094 | -0.0078 |
| | | 76.00 | 708.3843 | 708.3776 | 0.0067 | 29 | | 97.5 | 815.3192 | 815.3302 | -0.0110 |
| | | 76.50 | 711.2346 | 711.2272 | 0.0074 | | | 97.0 | 813.0361 | 813.0520 | -0.0159 |
| | | 77.00 | 714.0661 | 714.0604 | 0.0057 | | | 96.5 | 810.7518 | 810.7757 | -0.0239 |
| | 77.9 | 77.50 | 716.8776 | 716.8776 | 0.0000 | | | 96.0 | 808.4658 | 808.5028 | -0.0370 |
| | 77.9 | 77.90 | 719.1107 * | 719.1200 | -0.0093 | | | 95.618 | 806.7177 * | 806.7715 | -0.0538 |
| | | 78.50 | 722.4405 | 722.4426 | -0.0021 | | | 96.0 | 808.4658 | 808.4805 | -0.0147 |
| | | 79.00 | 725.1920 | 725.1893 | 0.0027 | | 96.289 | 96.289 | 809.7871 * | 809.7642 | 0.0229 |
| 21 | | 79.50 | 727.9234 | 727.9171 | 0.0063 | | | interpolated * | | | |
| | | 80.00 | 730.6349 | 730.6267 | 0.0082 | | | | | | |
| | | 80.50 | 733.3270 | 733.3190 | 0.0080 | | | | | | |
| | | 81.00 | 735.9998 | 735.9944 | 0.0054 | | | | | | |
| | 81.8 | 81.50 | 738.6533 | 738.6533 | 0.0000 | | | | | | |

| Table 2 | | | PcP travel times *(sec.)* | | | |
|---|---|---|---|---|---|---|
| Δ | Observed (Herrin & al.) | Computed MZY | Residual $T_o - T_c$ | Δ | Observed (Herrin & al.) | Computed MZY | Residual $T_o - T_c$ |
| 0 | 511.3 | 511.300 | 0.00 | 49 | 611.9 | 611.870 | 0.03 |
| 1 | 511.4 | 511.348 | 0.05 | 50 | 615.5 | 615.511 | -0.01 |
| 2 | 511.5 | 511.492 | 0.01 | 51 | 619.2 | 619.191 | 0.01 |
| 3 | 511.7 | 511.731 | -0.03 | 52 | 622.9 | 622.910 | -0.01 |
| 4 | 512.1 | 512.066 | 0.03 | 53 | 626.7 | 626.666 | 0.03 |
| 5 | 512.5 | 512.497 | 0.00 | 54 | 630.4 | 630.458 | -0.06 |
| 6 | 513.0 | 513.022 | -0.02 | 55 | 634.3 | 634.284 | 0.02 |
| 7 | 513.6 | 513.641 | -0.04 | 56 | 638.1 | 638.144 | -0.04 |
| 8 | 514.4 | 514.355 | 0.04 | 57 | 642.0 | 642.037 | -0.04 |
| 9 | 515.2 | 515.163 | 0.04 | 58 | 646.0 | 645.961 | 0.04 |
| 10 | 516.1 | 516.063 | 0.04 | 59 | 649.9 | 649.914 | -0.01 |
| 11 | 517.1 | 517.055 | 0.04 | 60 | 653.9 | 653.896 | 0.00 |
| 12 | 518.1 | 518.139 | -0.04 | 61 | 657.9 | 657.906 | -0.01 |
| 13 | 519.3 | 519.314 | -0.01 | 62 | 661.9 | 661.943 | -0.04 |
| 14 | 520.6 | 520.578 | 0.02 | 63 | 666.0 | 666.005 | 0.00 |
| 15 | 521.9 | 521.932 | -0.03 | 64 | 670.1 | 670.091 | 0.01 |
| 16 | 523.4 | 523.373 | 0.03 | 65 | 674.2 | 674.201 | 0.00 |
| 17 | 524.9 | 524.901 | 0.00 | 66 | 678.3 | 678.333 | -0.03 |
| 18 | 526.5 | 526.515 | -0.01 | 67 | 682.5 | 682.486 | 0.01 |
| 19 | 528.2 | 528.214 | -0.01 | 68 | 686.6 | 686.660 | -0.06 |
| 20 | 530.0 | 529.996 | 0.00 | 69 | 690.8 | 690.853 | -0.05 |
| 21 | 531.9 | 531.861 | 0.04 | 70 | 695.1 | 695.065 | 0.04 |
| 22 | 533.8 | 533.806 | -0.01 | 71 | 699.3 | 699.294 | 0.01 |
| 23 | 535.8 | 535.832 | -0.03 | 72 | 703.5 | 703.540 | -0.04 |
| 24 | 537.9 | 537.936 | -0.04 | 73 | 707.9 | 707.802 | 0.10 |
| 25 | 540.1 | 540.117 | -0.02 | 74 | 712.1 | 712.079 | 0.02 |
| 26 | 542.4 | 542.374 | 0.03 | 75 | 716.5 | 716.370 | 0.13 |
| 27 | 544.7 | 544.705 | -0.01 | 76 | 720.7 | 720.675 | 0.03 |
| 28 | 547.1 | 547.110 | -0.01 | 77 | 725.0 | 724.992 | 0.01 |
| 29 | 549.6 | 549.586 | 0.01 | 78 | 729.3 | 729.320 | -0.02 |
| 30 | 552.1 | 552.132 | -0.03 | 79 | 733.6 | 733.660 | -0.06 |
| 31 | 554.7 | 554.747 | -0.05 | 80 | 738.0 | 738.010 | -0.01 |
| 32 | 557.4 | 557.428 | -0.03 | 81 | 742.4 | 742.370 | 0.03 |
| 33 | 560.2 | 560.176 | 0.02 | 82 | 746.7 | 746.739 | -0.04 |
| 34 | 563.0 | 562.988 | 0.01 | 83 | 751.1 | 751.116 | -0.02 |
| 35 | 565.9 | 565.863 | 0.04 | 84 | 755.5 | 755.500 | 0.00 |
| 36 | 568.8 | 568.798 | 0.00 | 85 | 759.9 | 759.892 | 0.01 |
| 37 | 571.8 | 571.794 | 0.01 | 86 | 764.3 | 764.290 | 0.01 |
| 38 | 574.8 | 574.848 | -0.05 | 87 | 768.7 | 768.693 | 0.01 |
| 39 | 578.0 | 577.959 | 0.04 | 88 | 773.1 | 773.102 | 0.00 |
| 40 | 581.1 | 581.125 | -0.02 | 89 | 777.5 | 777.515 | -0.01 |
| 41 | 584.3 | 584.345 | -0.04 | 90 | 781.9 | 781.932 | -0.03 |
| 42 | 587.6 | 587.617 | -0.02 | 91 | 786.3 | 786.352 | -0.05 |
| 43 | 590.9 | 590.940 | -0.04 | 92 | 790.8 | 790.775 | 0.03 |
| 44 | 594.3 | 594.313 | -0.01 | 93 | 795.2 | 795.200 | 0.00 |
| 45 | 597.7 | 597.734 | -0.03 | 94 | 799.6 | 799.627 | -0.03 |
| 46 | 601.2 | 601.202 | 0.00 | 95 | 804.0 | 804.055 | -0.05 |
| 47 | 604.7 | 604.714 | -0.01 | 96 | 808.5 | 808.484 | 0.02 |
| 48 | 608.3 | 608.271 | 0.03 | 96.289 | | 809.764 | |

**Table 3**

| Data MZY Radius (km) Layers Depth (km) | $v_i$ (km/s) $B_i$ (×10⁻²) $A_i$ (×10⁻³) $v'_i$ (km/s) | Depth (km) | P-wave velocity (km/s) Radius of surface-focus = 6371.028 Herrin & al. | MZY | Residual H. & al. - MZY | Data MZY Radius (km) Layers Depth (km) | $v_i$ (km/s) $B_i$ (×10⁻²) $A_i$ (×10⁻³) $v'_i$ (km/s) | Depth (km) | P-wave velocity (km/s) Herrin & al. | MZY | Residual H. & al. - MZY |
|---|---|---|---|---|---|---|---|---|---|---|---|
| 6371.028 | 6.0000 | 0 | 6.0000 | 6.0001 | -0.0001 | 6021.290062 | 8.8862 | 350 | 8.8905 | 8.8875 | 0.0030 |
| **1** | 0.92341045 | 5 | 6.0000 | 6.0001 | -0.0001 | | | 355 | 8.9131 | 8.9114 | 0.0017 |
| | 0.94666548 | 10 | 6.0000 | 6.0001 | -0.0001 | | | 360 | 8.9360 | 8.9352 | 0.0008 |
| 15.001533 | 6.0001 | 15 | 6.0000 | 6.0001 | -0.0001 | | | 365 | 8.9590 | 8.9591 | -0.0001 |
| 6356.026467 | 6.7500 | 20 | 6.7500 | 6.7501 | -0.0001 | | | 370 | 8.9823 | 8.9829 | -0.0006 |
| **2** | 1.04491768 | 25 | 6.7500 | 6.7501 | -0.0001 | **10** | 5.59329351 | 375 | 9.0058 | 9.0067 | -0.0009 |
| | 1.07194400 | 30 | 6.7500 | 6.7502 | -0.0002 | | 6.25724227 | 380 | 9.0294 | 9.0304 | -0.0010 |
| | | 35 | 6.7500 | 6.7502 | -0.0002 | | | 385 | 9.0532 | 9.0542 | -0.0010 |
| 40.053935 | 6.7502 | 40 | 6.7500 | 6.7502 | -0.0002 | | | 390 | 9.0773 | 9.0779 | -0.0006 |
| 6330.974065 | 8.0540 | 45 | 8.0582 | 8.0593 | -0.0011 | | | 395 | 9.1015 | 9.1016 | -0.0001 |
| | | 50 | 8.0642 | 8.0645 | -0.0003 | | | 400 | 9.1258 | 9.1252 | 0.0006 |
| | | 55 | 8.0698 | 8.0698 | 0.0000 | | | 405 | 9.1503 | 9.1489 | 0.0014 |
| | | 60 | 8.0753 | 8.0751 | 0.0002 | 411.322422 | 9.1787 | 410 | 9.1750 | 9.1725 | 0.0025 |
| | | 65 | 8.0806 | 8.0803 | 0.0003 | 5959.705578 | 9.1787 | 415 | 9.1999 | 9.1977 | 0.0022 |
| **3** | 2.16670136 | 70 | 8.0859 | 8.0856 | 0.0003 | | | 420 | 9.2248 | 9.2236 | 0.0012 |
| | 2.32998530 | 75 | 8.0911 | 8.0908 | 0.0003 | | | 425 | 9.2499 | 9.2494 | 0.0005 |
| | | 80 | 8.0962 | 8.0960 | 0.0002 | | | 430 | 9.2752 | 9.2752 | 0.0000 |
| | | 85 | 8.1013 | 8.1012 | 0.0001 | | | 435 | 9.3007 | 9.3010 | -0.0003 |
| | | 90 | 8.1064 | 8.1064 | 0.0000 | **11** | 5.99325871 | 440 | 9.3262 | 9.3267 | -0.0005 |
| | | 95 | 8.1115 | 8.1116 | -0.0001 | | 6.71735442 | 445 | 9.3519 | 9.3524 | -0.0005 |
| 104.957687 | 8.1219 | 100 | 8.1165 | 8.1168 | -0.0003 | | | 450 | 9.3778 | 9.3781 | -0.0003 |
| 6266.070313 | 8.1219 | 105 | 8.1219 | 8.1220 | -0.0001 | | | 455 | 9.4038 | 9.4038 | 0.0000 |
| **4** | 2.51841026 | 110 | 8.1285 | 8.1292 | -0.0007 | | | 460 | 9.4299 | 9.4294 | 0.0005 |
| | 2.73226453 | 115 | 8.1356 | 8.1363 | -0.0007 | | | 465 | 9.4562 | 9.4550 | 0.0012 |
| | | 120 | 8.1432 | 8.1435 | -0.0003 | 472.071820 | 9.4911 | 470 | 9.4826 | 9.4805 | 0.0021 |
| 125.320 | 8.1511 | 125 | 8.1513 | 8.1506 | 0.0007 | 5898.956180 | 9.4911 | 475 | 9.5091 | 9.5072 | 0.0019 |
| 6245.707939 | 8.1511 | 130 | 8.1599 | 8.1602 | -0.0003 | | | 480 | 9.5358 | 9.5345 | 0.0013 |
| | | 135 | 8.1690 | 8.1699 | -0.0009 | | | 485 | 9.5626 | 9.5618 | 0.0008 |
| **5** | 2.97405843 | 140 | 8.1786 | 8.1796 | -0.0010 | | | 490 | 9.5895 | 9.5891 | 0.0004 |
| | 3.25362200 | 145 | 8.1886 | 8.1893 | -0.0007 | | | 495 | 9.6165 | 9.6164 | 0.0001 |
| | | 150 | 8.1991 | 8.1990 | 0.0001 | **12** | 6.31396167 | 500 | 9.6437 | 9.6436 | 0.0001 |
| 155.027311 | 8.2087 | 155 | 8.2101 | 8.2087 | 0.0014 | | 7.08672019 | 505 | 9.6709 | 9.6709 | 0.0000 |
| 6216.000689 | 8.2087 | 160 | 8.2214 | 8.2213 | 0.0001 | | | 510 | 9.6981 | 9.6980 | 0.0001 |
| | | 165 | 8.2332 | 8.2340 | -0.0008 | | | 515 | 9.7255 | 9.7252 | 0.0003 |
| **6** | 3.50244082 | 170 | 8.2454 | 8.2467 | -0.0013 | | | 520 | 9.7530 | 9.7523 | 0.0007 |
| | 3.85853274 | 175 | 8.2580 | 8.2593 | -0.0013 | | | 525 | 9.7805 | 9.7794 | 0.0011 |
| | | 180 | 8.2710 | 8.2719 | -0.0009 | 531.559687 | 9.8149 | 530 | 9.8080 | 9.8064 | 0.0016 |
| | | 185 | 8.2843 | 8.2845 | -0.0002 | 5839.468313 | 9.8149 | 535 | 9.8356 | 9.8342 | 0.0014 |
| 193.795314 | 8.3067 | 190 | 8.2980 | 8.2971 | 0.0009 | | | 540 | 9.8632 | 9.8623 | 0.0009 |
| 6177.232686 | 8.3067 | 195 | 8.3120 | 8.3104 | 0.0016 | | | 545 | 9.8908 | 9.8903 | 0.0005 |
| | | 200 | 8.3264 | 8.3260 | 0.0004 | | | 550 | 9.9185 | 9.9184 | 0.0001 |
| | | 205 | 8.3410 | 8.3415 | -0.0005 | | | 555 | 9.9462 | 9.9464 | -0.0002 |
| **7** | 4.03019188 | 210 | 8.3560 | 8.3571 | -0.0011 | | | 560 | 9.9740 | 9.9743 | -0.0003 |
| | 4.46315376 | 215 | 8.3713 | 8.3726 | -0.0013 | | | 565 | 10.0018 | 10.0023 | -0.0005 |
| | | 220 | 8.3870 | 8.3881 | -0.0011 | | | 570 | 10.0296 | 10.0302 | -0.0006 |
| | | 225 | 8.4029 | 8.4036 | -0.0007 | | | 575 | 10.0574 | 10.0580 | -0.0006 |
| | | 230 | 8.4191 | 8.4191 | 0.0000 | | | 580 | 10.0853 | 10.0859 | -0.0006 |
| | | 235 | 8.4357 | 8.4345 | 0.0012 | | | 585 | 10.1132 | 10.1137 | -0.0005 |
| 239.227039 | 8.4476 | 240 | 8.4525 | 8.4504 | 0.0021 | | | 590 | 10.1411 | 10.1414 | -0.0003 |
| 6131.800961 | 8.4476 | 245 | 8.4696 | 8.4689 | 0.0007 | | | 595 | 10.1690 | 10.1692 | -0.0002 |
| | | 250 | 8.4870 | 8.4873 | -0.0003 | | | 600 | 10.1970 | 10.1969 | 0.0001 |
| | | 255 | 8.5047 | 8.5057 | -0.0010 | | | 605 | 10.2249 | 10.2246 | 0.0003 |
| **8** | 4.55770255 | 260 | 8.5227 | 8.5241 | -0.0014 | **13** | 6.50401417 | 610 | 10.2528 | 10.2522 | 0.0006 |
| | 5.06801092 | 265 | 8.5410 | 8.5424 | -0.0014 | | 7.30586668 | 615 | 10.2807 | 10.2798 | 0.0009 |
| | | 270 | 8.5595 | 8.5607 | -0.0012 | | | 620 | 10.3086 | 10.3074 | 0.0012 |
| | | 275 | 8.5783 | 8.5791 | -0.0008 | | | 625 | 10.3364 | 10.3350 | 0.0014 |
| | | 280 | 8.5973 | 8.5974 | -0.0001 | | | 630 | 10.3642 | 10.3625 | 0.0017 |
| | | 285 | 8.6167 | 8.6156 | 0.0011 | | | 635 | 10.3920 | 10.3900 | 0.0020 |
| 291.392218 | 8.6390 | 290 | 8.6362 | 8.6339 | 0.0023 | | | 640 | 10.4197 | 10.4174 | 0.0023 |
| 6079.635782 | 8.6390 | 295 | 8.6561 | 8.6543 | 0.0018 | | | 645 | 10.4474 | 10.4449 | 0.0025 |
| | | 300 | 8.6762 | 8.6756 | 0.0006 | | | 650 | 10.4750 | 10.4723 | 0.0027 |
| | | 305 | 8.6966 | 8.6969 | -0.0003 | | | 655 | 10.5024 | 10.4996 | 0.0028 |
| | | 310 | 8.7172 | 8.7182 | -0.0010 | | | 660 | 10.5297 | 10.5270 | 0.0027 |
| **9** | 5.09632938 | 315 | 8.7380 | 8.7394 | -0.0014 | | | 665 | 10.5570 | 10.5542 | 0.0028 |
| | 5.68621977 | 320 | 8.7591 | 8.7606 | -0.0015 | | | 670 | 10.5840 | 10.5815 | 0.0025 |
| | | 325 | 8.7804 | 8.7818 | -0.0014 | | | 675 | 10.6109 | 10.6087 | 0.0022 |
| | | 330 | 8.8020 | 8.8029 | -0.0009 | | | 680 | 10.6375 | 10.6359 | 0.0016 |
| | | 335 | 8.8238 | 8.8241 | -0.0003 | | | 685 | 10.6638 | 10.6631 | 0.0007 |
| | | 340 | 8.8458 | 8.8452 | 0.0006 | | | 690 | 10.6899 | 10.6903 | -0.0004 |
| 349.737938 | 8.8862 | 345 | 8.8680 | 8.8663 | 0.0017 | | | 695 | 10.7157 | 10.7174 | -0.0017 |
| | | | | | | 700.177664 | 10.7454 | 700 | 10.7412 | 10.7444 | -0.0032 |

p. 1/4

**Table 3**

| Data MZY | | | P-wave velocity (km/s) | | | Data MZY | | p. 2/4 | P-wave velocity (km/s) | | |
|---|---|---|---|---|---|---|---|---|---|---|---|
| Radius (km) | $v_i$ (km/s) | Depth (km) | Radius of surface-focus = 6371.028 | | | Radius (km) | $v_i$ (km/s) | Depth (km) | | | |
| **Layers** | $B_i$ ( x $10^{-2}$ ) | | | | | **Layers** | $B_i$ ( x $10^{-2}$ ) | | | | |
| | $A_i$ ( x $10^{-3}$ ) | | Herrin & al. | MZY | Residual | | $A_i$ ( x $10^{-3}$ ) | | Herrin & al. | MZY | Residual |
| Depth (km) | $v'_i$ (km/s) | | | | H. & al. - MZY | Depth (km) | $v'_i$ (km/s) | | | | H. & al. - MZY |
| 5670.850336 | 10.7454 | 705 | 10.7664 | 10.7679 | -0.0015 | | | 1115 | 11.6288 | 11.6296 | -0.0008 |
| | | 710 | 10.7911 | 10.7912 | -0.0001 | | | 1120 | 11.6367 | 11.6375 | -0.0008 |
| | | 715 | 10.8154 | 10.8144 | 0.0010 | | | 1125 | 11.6446 | 11.6455 | -0.0009 |
| 14 | 5.85891401 | 720 | 10.8392 | 10.8376 | 0.0016 | | | 1130 | 11.6525 | 11.6534 | -0.0009 |
| | 6.55949029 | 725 | 10.8624 | 10.8608 | 0.0016 | | | 1135 | 11.6604 | 11.6613 | -0.0009 |
| | | 730 | 10.8850 | 10.8840 | 0.0010 | | | 1140 | 11.6684 | 11.6692 | -0.0008 |
| | | 735 | 10.9068 | 10.9071 | -0.0003 | | | 1145 | 11.6763 | 11.6771 | -0.0008 |
| 744.109784 | 10.9492 | 740 | 10.9279 | 10.9302 | -0.0023 | | | 1150 | 11.6842 | 11.6849 | -0.0007 |
| 5626.918216 | 10.9492 | 745 | 10.9479 | 10.9515 | -0.0036 | | | 1155 | 11.6921 | 11.6928 | -0.0007 |
| | | 750 | 10.9663 | 10.9645 | 0.0018 | | | 1160 | 11.7000 | 11.7006 | -0.0006 |
| 15 | 4.12047826 | 755 | 10.9819 | 10.9775 | 0.0044 | | | 1165 | 11.7080 | 11.7084 | -0.0004 |
| | 4.54632080 | 760 | 10.9933 | 10.9904 | 0.0029 | | | 1170 | 11.7159 | 11.7162 | -0.0003 |
| | | 765 | 11.0029 | 11.0033 | -0.0004 | | | 1175 | 11.7238 | 11.7239 | -0.0001 |
| 772.052042 | 11.0215 | 770 | 11.0134 | 11.0163 | -0.0029 | | | 1180 | 11.7316 | 11.7317 | -0.0001 |
| 5598.975958 | 11.0215 | 775 | 11.0240 | 11.0273 | -0.0033 | | | 1185 | 11.7395 | 11.7394 | 0.0001 |
| | | 780 | 11.0348 | 11.0370 | -0.0022 | | | 1190 | 11.7473 | 11.7471 | 0.0002 |
| | | 785 | 11.0455 | 11.0467 | -0.0012 | | | 1195 | 11.7551 | 11.7548 | 0.0003 |
| | | 790 | 11.0561 | 11.0564 | -0.0003 | | | 1200 | 11.7629 | 11.7624 | 0.0005 |
| | | 795 | 11.0666 | 11.0660 | 0.0006 | | | 1205 | 11.7707 | 11.7701 | 0.0006 |
| | | 800 | 11.0766 | 11.0757 | 0.0009 | | | 1210 | 11.7785 | 11.7777 | 0.0008 |
| | | 805 | 11.0865 | 11.0853 | 0.0012 | | | 1215 | 11.7862 | 11.7853 | 0.0009 |
| | | 810 | 11.0962 | 11.0949 | 0.0013 | | | 1220 | 11.7938 | 11.7929 | 0.0009 |
| | | 815 | 11.1060 | 11.1045 | 0.0015 | | | 1225 | 11.8015 | 11.8005 | 0.0010 |
| | | 820 | 11.1156 | 11.1141 | 0.0015 | | | 1230 | 11.8091 | 11.8081 | 0.0010 |
| | | 825 | 11.1252 | 11.1236 | 0.0016 | | | 1235 | 11.8167 | 11.8156 | 0.0011 |
| | | 830 | 11.1348 | 11.1332 | 0.0016 | | | 1240 | 11.8242 | 11.8231 | 0.0011 |
| | | 835 | 11.1443 | 11.1427 | 0.0016 | | | 1245 | 11.8318 | 11.8306 | 0.0012 |
| | | 840 | 11.1538 | 11.1522 | 0.0016 | | | 1250 | 11.8393 | 11.8381 | 0.0012 |
| | | 845 | 11.1632 | 11.1617 | 0.0015 | | | 1255 | 11.8467 | 11.8456 | 0.0011 |
| | | 850 | 11.1726 | 11.1711 | 0.0015 | | | 1260 | 11.8542 | 11.8530 | 0.0012 |
| 16 | 3.57557893 | 855 | 11.1819 | 11.1806 | 0.0013 | | | 1265 | 11.8616 | 11.8605 | 0.0011 |
| | 3.91494425 | 860 | 11.1912 | 11.1900 | 0.0012 | | | 1270 | 11.8689 | 11.8679 | 0.0010 |
| | | 865 | 11.2004 | 11.1994 | 0.0010 | | | 1275 | 11.8763 | 11.8753 | 0.0010 |
| | | 870 | 11.2096 | 11.2088 | 0.0008 | | | 1280 | 11.8836 | 11.8826 | 0.0010 |
| | | 875 | 11.2189 | 11.2182 | 0.0007 | | | 1285 | 11.8909 | 11.8900 | 0.0009 |
| | | 880 | 11.2281 | 11.2276 | 0.0005 | | | 1290 | 11.8982 | 11.8973 | 0.0009 |
| | | 885 | 11.2373 | 11.2369 | 0.0004 | | | 1295 | 11.9054 | 11.9046 | 0.0008 |
| | | 890 | 11.2466 | 11.2462 | 0.0004 | 17 | 3.48245516 | 1300 | 11.9126 | 11.9119 | 0.0007 |
| | | 895 | 11.2558 | 11.2555 | 0.0003 | | 3.80663410 | 1305 | 11.9198 | 11.9192 | 0.0006 |
| | | 900 | 11.2651 | 11.2648 | 0.0003 | | | 1310 | 11.9269 | 11.9265 | 0.0004 |
| | | 905 | 11.2743 | 11.2741 | 0.0002 | | | 1315 | 11.9341 | 11.9337 | 0.0004 |
| | | 910 | 11.2835 | 11.2833 | 0.0002 | | | 1320 | 11.9412 | 11.9409 | 0.0003 |
| | | 915 | 11.2927 | 11.2926 | 0.0001 | | | 1325 | 11.9483 | 11.9481 | 0.0002 |
| | | 920 | 11.3019 | 11.3018 | 0.0001 | | | 1330 | 11.9554 | 11.9553 | 0.0001 |
| | | 925 | 11.3110 | 11.3110 | 0.0000 | | | 1335 | 11.9624 | 11.9625 | -0.0001 |
| | | 930 | 11.3201 | 11.3202 | -0.0001 | | | 1340 | 11.9695 | 11.9696 | -0.0001 |
| | | 935 | 11.3291 | 11.3293 | -0.0002 | | | 1345 | 11.9765 | 11.9768 | -0.0003 |
| | | 940 | 11.3381 | 11.3385 | -0.0004 | | | 1350 | 11.9836 | 11.9839 | -0.0003 |
| | | 945 | 11.3470 | 11.3476 | -0.0006 | | | 1355 | 11.9906 | 11.9910 | -0.0004 |
| 950.864911 | 11.3583 | 950 | 11.3558 | 11.3567 | -0.0009 | | | 1360 | 11.9976 | 11.9980 | -0.0004 |
| 5420.163089 | 11.3583 | 955 | 11.3646 | 11.3654 | -0.0008 | | | 1365 | 12.0046 | 12.0051 | -0.0005 |
| | | 960 | 11.3734 | 11.3739 | -0.0005 | | | 1370 | 12.0116 | 12.0121 | -0.0005 |
| | | 965 | 11.3820 | 11.3824 | -0.0004 | | | 1375 | 12.0185 | 12.0191 | -0.0006 |
| | | 970 | 11.3907 | 11.3909 | -0.0002 | | | 1380 | 12.0255 | 12.0261 | -0.0006 |
| | | 975 | 11.3993 | 11.3994 | -0.0001 | | | 1385 | 12.0324 | 12.0331 | -0.0007 |
| | | 980 | 11.4079 | 11.4079 | 0.0000 | | | 1390 | 12.0393 | 12.0401 | -0.0008 |
| | | 985 | 11.4164 | 11.4163 | 0.0001 | | | 1395 | 12.0462 | 12.0470 | -0.0008 |
| | | 990 | 11.4249 | 11.4247 | 0.0002 | | | 1400 | 12.0531 | 12.0539 | -0.0008 |
| | | 995 | 11.4333 | 11.4331 | 0.0002 | | | 1405 | 12.0599 | 12.0608 | -0.0009 |
| | | 1000 | 11.4418 | 11.4415 | 0.0003 | | | 1410 | 12.0668 | 12.0677 | -0.0009 |
| | | 1005 | 11.4502 | 11.4499 | 0.0003 | | | 1415 | 12.0736 | 12.0746 | -0.0010 |
| | | 1010 | 11.4585 | 11.4582 | 0.0003 | | | 1420 | 12.0805 | 12.0814 | -0.0009 |
| | | 1015 | 11.4668 | 11.4666 | 0.0002 | | | 1425 | 12.0873 | 12.0882 | -0.0009 |
| | | 1020 | 11.4751 | 11.4749 | 0.0002 | | | 1430 | 12.0941 | 12.0950 | -0.0009 |
| | | 1025 | 11.4834 | 11.4832 | 0.0002 | | | 1435 | 12.1009 | 12.1018 | -0.0009 |
| | | 1030 | 11.4917 | 11.4915 | 0.0002 | | | 1440 | 12.1077 | 12.1086 | -0.0009 |
| | | 1035 | 11.4999 | 11.4998 | 0.0001 | | | 1445 | 12.1145 | 12.1153 | -0.0008 |
| 17 | 3.48245516 | 1040 | 11.5081 | 11.5080 | 0.0001 | | | 1450 | 12.1213 | 12.1221 | -0.0008 |
| | 3.80663410 | 1045 | 11.5163 | 11.5162 | 0.0001 | | | 1455 | 12.1281 | 12.1288 | -0.0007 |
| | | 1050 | 11.5244 | 11.5245 | -0.0001 | | | 1460 | 12.1349 | 12.1354 | -0.0005 |
| | | 1055 | 11.5326 | 11.5326 | 0.0000 | | | 1465 | 12.1417 | 12.1421 | -0.0004 |
| | | 1060 | 11.5407 | 11.5408 | -0.0001 | | | 1470 | 12.1485 | 12.1488 | -0.0003 |
| | | 1065 | 11.5488 | 11.5490 | -0.0002 | | | 1475 | 12.1553 | 12.1554 | -0.0001 |
| | | 1070 | 11.5569 | 11.5571 | -0.0002 | | | 1480 | 12.1620 | 12.1620 | 0.0000 |
| | | 1075 | 11.5649 | 11.5652 | -0.0003 | | | 1485 | 12.1688 | 12.1686 | 0.0002 |
| | | 1080 | 11.5730 | 11.5734 | -0.0004 | | | 1490 | 12.1755 | 12.1752 | 0.0003 |
| | | 1085 | 11.5810 | 11.5814 | -0.0004 | | | 1495 | 12.1822 | 12.1817 | 0.0005 |
| | | 1090 | 11.5890 | 11.5895 | -0.0005 | | | 1500 | 12.1889 | 12.1882 | 0.0007 |
| | | 1095 | 11.5970 | 11.5976 | -0.0006 | | | 1505 | 12.1956 | 12.1948 | 0.0008 |
| | | 1100 | 11.6049 | 11.6056 | -0.0007 | | | 1510 | 12.2023 | 12.2013 | 0.0010 |
| | | 1105 | 11.6129 | 11.6136 | -0.0007 | 1516.957451 | 12.2103 | 1515 | 12.2089 | 12.2077 | 0.0012 |
| | | 1110 | 11.6208 | 11.6216 | -0.0008 | | | | | | |

Table 3

| Data MZY | | P-wave velocity (km/s) | | | Data MZY | | p. 3/4 | P-wave velocity (km/s) | | |
|---|---|---|---|---|---|---|---|---|---|---|
| Radius (km) Layers Depth (km) | $v_i$ (km/s) $B_i$ (× 10$^{-2}$) $A_i$ (× 10$^{-3}$) $v'_i$ (km/s) | Depth (km) | Radius of surface-focus = 6371.028 | | | Radius (km) Layers Depth (km) | $v_i$ (km/s) $B_i$ (× 10$^{-2}$) $A_i$ (× 10$^{-3}$) $v'_i$ (km/s) | Depth (km) | | | |
| | | | Herrin & al. | MZY | Residual H. & al. - MZY | | | | Herrin & al. | MZY | Residual H. & al. - MZY |
| 4854.070549 | 12.2103 | 1520 | 12.2155 | 12.2142 | 0.0013 | 4526.927109 | 12.5946 | 1845 | 12.5982 | 12.5957 | 0.0025 |
| | | 1525 | 12.2221 | 12.2207 | 0.0014 | | | 1850 | 12.6037 | 12.6018 | 0.0019 |
| | | 1530 | 12.2287 | 12.2272 | 0.0015 | | | 1855 | 12.6093 | 12.6079 | 0.0014 |
| | | 1535 | 12.2352 | 12.2337 | 0.0015 | | | 1860 | 12.6149 | 12.6141 | 0.0008 |
| | | 1540 | 12.2417 | 12.2402 | 0.0015 | | | 1865 | 12.6205 | 12.6201 | 0.0004 |
| | | 1545 | 12.2482 | 12.2466 | 0.0016 | | | 1870 | 12.6262 | 12.6262 | 0.0000 |
| | | 1550 | 12.2547 | 12.2530 | 0.0017 | | | 1875 | 12.6319 | 12.6322 | -0.0003 |
| | | 1555 | 12.2611 | 12.2594 | 0.0017 | | | 1880 | 12.6376 | 12.6383 | -0.0007 |
| | | 1560 | 12.2675 | 12.2658 | 0.0017 | | | 1885 | 12.6433 | 12.6443 | -0.0010 |
| | | 1565 | 12.2738 | 12.2721 | 0.0017 | | | 1890 | 12.6491 | 12.6502 | -0.0011 |
| | | 1570 | 12.2801 | 12.2784 | 0.0017 | | | 1895 | 12.6549 | 12.6562 | -0.0013 |
| | | 1575 | 12.2864 | 12.2848 | 0.0016 | | | 1900 | 12.6607 | 12.6621 | -0.0014 |
| | | 1580 | 12.2926 | 12.2910 | 0.0016 | 19 | 3.65834366 | 1905 | 12.6666 | 12.6680 | -0.0014 |
| | | 1585 | 12.2988 | 12.2973 | 0.0015 | | 4.01545527 | 1910 | 12.6725 | 12.6739 | -0.0014 |
| | | 1590 | 12.3050 | 12.3036 | 0.0014 | | | 1915 | 12.6784 | 12.6798 | -0.0014 |
| | | 1595 | 12.3111 | 12.3098 | 0.0013 | | | 1920 | 12.6843 | 12.6856 | -0.0013 |
| | | 1600 | 12.3172 | 12.3160 | 0.0012 | | | 1925 | 12.6902 | 12.6914 | -0.0012 |
| | | 1605 | 12.3232 | 12.3222 | 0.0010 | | | 1930 | 12.6962 | 12.6972 | -0.0010 |
| | | 1610 | 12.3292 | 12.3284 | 0.0008 | | | 1935 | 12.7022 | 12.7030 | -0.0008 |
| | | 1615 | 12.3352 | 12.3345 | 0.0007 | | | 1940 | 12.7082 | 12.7087 | -0.0005 |
| | | 1620 | 12.3411 | 12.3407 | 0.0004 | | | 1945 | 12.7142 | 12.7144 | -0.0002 |
| | | 1625 | 12.3471 | 12.3468 | 0.0003 | | | 1950 | 12.7202 | 12.7202 | 0.0000 |
| | | 1630 | 12.3530 | 12.3529 | 0.0001 | | | 1955 | 12.7262 | 12.7258 | 0.0004 |
| | | 1635 | 12.3589 | 12.3589 | 0.0000 | | | 1960 | 12.7323 | 12.7315 | 0.0008 |
| | | 1640 | 12.3648 | 12.3650 | -0.0002 | | | 1965 | 12.7383 | 12.7371 | 0.0012 |
| | | 1645 | 12.3706 | 12.3710 | -0.0004 | | | 1970 | 12.7444 | 12.7427 | 0.0017 |
| | | 1650 | 12.3765 | 12.3770 | -0.0005 | | | 1975 | 12.7505 | 12.7483 | 0.0022 |
| | | 1655 | 12.3823 | 12.3830 | -0.0007 | 1981.009473 | 12.7550 | 1980 | 12.7565 | 12.7539 | 0.0026 |
| | | 1660 | 12.3882 | 12.3890 | -0.0008 | 4390.018527 | 12.7550 | 1985 | 12.7626 | 12.7601 | 0.0025 |
| | | 1665 | 12.3940 | 12.3950 | -0.0010 | | | 1990 | 12.7687 | 12.7664 | 0.0023 |
| 18 | 3.49547729 | 1670 | 12.3999 | 12.4009 | -0.0010 | | | 1995 | 12.7747 | 12.7726 | 0.0021 |
| | 3.82197670 | 1675 | 12.4057 | 12.4068 | -0.0011 | | | 2000 | 12.7808 | 12.7789 | 0.0019 |
| | | 1680 | 12.4115 | 12.4127 | -0.0012 | | | 2005 | 12.7868 | 12.7851 | 0.0017 |
| | | 1685 | 12.4173 | 12.4186 | -0.0013 | | | 2010 | 12.7928 | 12.7913 | 0.0015 |
| | | 1690 | 12.4231 | 12.4244 | -0.0013 | | | 2015 | 12.7988 | 12.7975 | 0.0013 |
| | | 1695 | 12.4289 | 12.4302 | -0.0013 | | | 2020 | 12.8048 | 12.8037 | 0.0011 |
| | | 1700 | 12.4347 | 12.4360 | -0.0013 | | | 2025 | 12.8108 | 12.8098 | 0.0010 |
| | | 1705 | 12.4405 | 12.4418 | -0.0013 | | | 2030 | 12.8167 | 12.8159 | 0.0008 |
| | | 1710 | 12.4463 | 12.4476 | -0.0013 | | | 2035 | 12.8227 | 12.8220 | 0.0007 |
| | | 1715 | 12.4521 | 12.4533 | -0.0012 | | | 2040 | 12.8286 | 12.8280 | 0.0006 |
| | | 1720 | 12.4578 | 12.4591 | -0.0013 | | | 2045 | 12.8345 | 12.8341 | 0.0004 |
| | | 1725 | 12.4636 | 12.4648 | -0.0012 | | | 2050 | 12.8403 | 12.8401 | 0.0002 |
| | | 1730 | 12.4693 | 12.4705 | -0.0012 | | | 2055 | 12.8462 | 12.8461 | 0.0001 |
| | | 1735 | 12.4751 | 12.4761 | -0.0010 | | | 2060 | 12.8520 | 12.8520 | 0.0000 |
| | | 1740 | 12.4808 | 12.4818 | -0.0010 | | | 2065 | 12.8578 | 12.8580 | -0.0002 |
| | | 1745 | 12.4865 | 12.4874 | -0.0009 | | | 2070 | 12.8636 | 12.8639 | -0.0003 |
| | | 1750 | 12.4922 | 12.4930 | -0.0008 | | | 2075 | 12.8693 | 12.8698 | -0.0005 |
| | | 1755 | 12.4978 | 12.4986 | -0.0008 | 20 | 3.78936074 | 2080 | 12.8751 | 12.8757 | -0.0006 |
| | | 1760 | 12.5035 | 12.5041 | -0.0006 | | 4.17166810 | 2085 | 12.8808 | 12.8815 | -0.0007 |
| | | 1765 | 12.5091 | 12.5097 | -0.0006 | | | 2090 | 12.8865 | 12.8873 | -0.0008 |
| | | 1770 | 12.5148 | 12.5152 | -0.0004 | | | 2095 | 12.8922 | 12.8931 | -0.0009 |
| | | 1775 | 12.5204 | 12.5207 | -0.0003 | | | 2100 | 12.8979 | 12.8989 | -0.0010 |
| | | 1780 | 12.5259 | 12.5262 | -0.0003 | | | 2105 | 12.9036 | 12.9046 | -0.0010 |
| | | 1785 | 12.5315 | 12.5316 | -0.0001 | | | 2110 | 12.9092 | 12.9104 | -0.0012 |
| | | 1790 | 12.5371 | 12.5371 | 0.0000 | | | 2115 | 12.9149 | 12.9161 | -0.0012 |
| | | 1795 | 12.5426 | 12.5425 | 0.0001 | | | 2120 | 12.9205 | 12.9217 | -0.0012 |
| | | 1800 | 12.5481 | 12.5479 | 0.0002 | | | 2125 | 12.9262 | 12.9274 | -0.0012 |
| | | 1805 | 12.5537 | 12.5533 | 0.0004 | | | 2130 | 12.9318 | 12.9330 | -0.0012 |
| | | 1810 | 12.5592 | 12.5586 | 0.0006 | | | 2135 | 12.9375 | 12.9386 | -0.0011 |
| | | 1815 | 12.5648 | 12.5639 | 0.0009 | | | 2140 | 12.9431 | 12.9442 | -0.0011 |
| | | 1820 | 12.5703 | 12.5693 | 0.0010 | | | 2145 | 12.9487 | 12.9497 | -0.0010 |
| | | 1825 | 12.5759 | 12.5745 | 0.0014 | | | 2150 | 12.9544 | 12.9552 | -0.0008 |
| | | 1830 | 12.5814 | 12.5798 | 0.0016 | | | 2155 | 12.9601 | 12.9607 | -0.0006 |
| | | 1835 | 12.5870 | 12.5851 | 0.0019 | | | 2160 | 12.9657 | 12.9662 | -0.0005 |
| 1844.100891 | 12.5946 | 1840 | 12.5926 | 12.5903 | 0.0023 | | | 2165 | 12.9714 | 12.9717 | -0.0003 |
| | | | | | | | | 2170 | 12.9771 | 12.9771 | 0.0000 |
| | | | | | | | | 2175 | 12.9829 | 12.9825 | 0.0004 |
| | | | | | | | | 2180 | 12.9886 | 12.9879 | 0.0007 |
| | | | | | | | | 2185 | 12.9944 | 12.9932 | 0.0012 |
| | | | | | | | | 2190 | 13.0002 | 12.9985 | 0.0017 |
| | | | | | | | | 2195 | 13.0059 | 13.0038 | 0.0021 |
| | | | | | | 2201.417284 | 13.0106 | 2200 | 13.0117 | 13.0091 | 0.0026 |

Table 3

| Data MZY | | | P-wave velocity (km/s) | | | Data MZY | | p. 4/4 | P-wave velocity (km/s) | | |
|---|---|---|---|---|---|---|---|---|---|---|---|
| Radius (km) | $v_i$ (km/s) | | Radius of surface-focus = 6371.028 | | | Radius (km) | $v_i$ (km/s) | | | | |
| **Layers** | $B_i$ ( x $10^{-2}$ ) | Depth | | | | **Layers** | $B_i$ ( x $10^{-2}$ ) | Depth | | | |
| | $A_i$ ( x $10^{-3}$ ) | (km) | Herrin & al. | MZY | Residual | | $A_i$ ( x $10^{-3}$ ) | (km) | Herrin & al. | MZY | Residual |
| Depth (km) | $v'_i$ (km/s) | | | | H. & al. - MZY | Depth (km) | $v'_i$ (km/s) | | | | H. & al. - MZY |
| 4169.610716 | 13.0106 | 2205 | 13.0175 | 13.0151 | 0.0024 | 3887.018610 | 13.3469 | 2485 | 13.3499 | 13.3482 | 0.0017 |
| | | 2210 | 13.0234 | 13.0214 | 0.0020 | | | 2490 | 13.3562 | 13.3549 | 0.0013 |
| | | 2215 | 13.0292 | 13.0277 | 0.0015 | | | 2495 | 13.3626 | 13.3615 | 0.0011 |
| | | 2220 | 13.0350 | 13.0339 | 0.0011 | | | 2500 | 13.3690 | 13.3680 | 0.0010 |
| | | 2225 | 13.0408 | 13.0401 | 0.0007 | | | 2505 | 13.3753 | 13.3746 | 0.0007 |
| | | 2230 | 13.0466 | 13.0462 | 0.0004 | | | 2510 | 13.3817 | 13.3811 | 0.0006 |
| | | 2235 | 13.0525 | 13.0524 | 0.0001 | | | 2515 | 13.3881 | 13.3876 | 0.0005 |
| | | 2240 | 13.0583 | 13.0585 | -0.0002 | | | 2520 | 13.3945 | 13.3940 | 0.0005 |
| | | 2245 | 13.0641 | 13.0646 | -0.0005 | | | 2525 | 13.4009 | 13.4004 | 0.0005 |
| | | 2250 | 13.0700 | 13.0707 | -0.0007 | | | 2530 | 13.4073 | 13.4068 | 0.0005 |
| | | 2255 | 13.0758 | 13.0767 | -0.0009 | | | 2535 | 13.4137 | 13.4132 | 0.0005 |
| | | 2260 | 13.0817 | 13.0827 | -0.0010 | | | 2540 | 13.4200 | 13.4195 | 0.0005 |
| | | 2265 | 13.0876 | 13.0887 | -0.0011 | | | 2545 | 13.4264 | 13.4258 | 0.0006 |
| | | 2270 | 13.0934 | 13.0947 | -0.0013 | | | 2550 | 13.4327 | 13.4321 | 0.0006 |
| | | 2275 | 13.0993 | 13.1006 | -0.0013 | | | 2555 | 13.4391 | 13.4383 | 0.0008 |
| **21** | 3.96471462 | 2280 | 13.1052 | 13.1065 | -0.0013 | | | 2560 | 13.4454 | 13.4445 | 0.0009 |
| | 4.38203609 | 2285 | 13.1110 | 13.1124 | -0.0014 | **23** | 4.28150596 | 2565 | 13.4516 | 13.4507 | 0.0009 |
| | | 2290 | 13.1169 | 13.1182 | -0.0013 | | 4.76460375 | 2570 | 13.4578 | 13.4568 | 0.0010 |
| | | 2295 | 13.1228 | 13.1241 | -0.0013 | | | 2575 | 13.4640 | 13.4629 | 0.0011 |
| | | 2300 | 13.1287 | 13.1298 | -0.0011 | | | 2580 | 13.4702 | 13.4690 | 0.0012 |
| | | 2305 | 13.1345 | 13.1356 | -0.0011 | | | 2585 | 13.4763 | 13.4750 | 0.0013 |
| | | 2310 | 13.1404 | 13.1414 | -0.0010 | | | 2590 | 13.4823 | 13.4810 | 0.0013 |
| | | 2315 | 13.1462 | 13.1471 | -0.0009 | | | 2595 | 13.4883 | 13.4870 | 0.0013 |
| | | 2320 | 13.1521 | 13.1528 | -0.0007 | | | 2600 | 13.4942 | 13.4930 | 0.0012 |
| | | 2325 | 13.1579 | 13.1584 | -0.0005 | | | 2605 | 13.5000 | 13.4989 | 0.0011 |
| | | 2330 | 13.1637 | 13.1641 | -0.0004 | | | 2610 | 13.5058 | 13.5048 | 0.0010 |
| | | 2335 | 13.1696 | 13.1697 | -0.0001 | | | 2615 | 13.5116 | 13.5106 | 0.0010 |
| | | 2340 | 13.1754 | 13.1753 | 0.0001 | | | 2620 | 13.5172 | 13.5164 | 0.0008 |
| | | 2345 | 13.1812 | 13.1808 | 0.0004 | | | 2625 | 13.5229 | 13.5222 | 0.0007 |
| | | 2350 | 13.1871 | 13.1863 | 0.0008 | | | 2630 | 13.5285 | 13.5280 | 0.0005 |
| | | 2355 | 13.1929 | 13.1918 | 0.0011 | | | 2635 | 13.5340 | 13.5337 | 0.0003 |
| | | 2360 | 13.1987 | 13.1973 | 0.0014 | | | 2640 | 13.5394 | 13.5394 | 0.0000 |
| | | 2365 | 13.2046 | 13.2028 | 0.0018 | | | 2645 | 13.5448 | 13.5451 | -0.0003 |
| 2372.501241 | 13.2109 | 2370 | 13.2104 | 13.2082 | 0.0022 | | | 2650 | 13.5501 | 13.5507 | -0.0006 |
| 3998.526759 | 13.2109 | 2375 | 13.2163 | 13.2141 | 0.0022 | | | 2655 | 13.5554 | 13.5563 | -0.0009 |
| | | 2380 | 13.2221 | 13.2205 | 0.0016 | 2660.414706 | 13.5623 | 2660 | 13.5606 | 13.5619 | -0.0013 |
| | | 2385 | 13.2280 | 13.2268 | 0.0012 | 3710.613294 | 13.5623 | 2665 | 13.5657 | 13.5666 | -0.0009 |
| | | 2390 | 13.2339 | 13.2332 | 0.0007 | | | 2670 | 13.5707 | 13.5712 | -0.0005 |
| | | 2395 | 13.2397 | 13.2395 | 0.0002 | | | 2675 | 13.5756 | 13.5758 | -0.0002 |
| | | 2400 | 13.2456 | 13.2458 | -0.0002 | | | 2680 | 13.5804 | 13.5804 | 0.0000 |
| | | 2405 | 13.2516 | 13.2520 | -0.0004 | **24** | 4.13634936 | 2685 | 13.5851 | 13.5849 | 0.0002 |
| | | 2410 | 13.2575 | 13.2582 | -0.0007 | | 4.58799171 | 2690 | 13.5898 | 13.5894 | 0.0004 |
| | | 2415 | 13.2635 | 13.2644 | -0.0009 | | | 2695 | 13.5942 | 13.5938 | 0.0004 |
| **22** | 4.13594750 | 2420 | 13.2694 | 13.2706 | -0.0012 | | | 2700 | 13.5986 | 13.5983 | 0.0003 |
| | 4.58849794 | 2425 | 13.2755 | 13.2767 | -0.0012 | | | 2705 | 13.6027 | 13.6027 | 0.0000 |
| | | 2430 | 13.2815 | 13.2828 | -0.0013 | 2711.399520 | 13.6083 | 2710 | 13.6067 | 13.6071 | -0.0004 |
| | | 2435 | 13.2876 | 13.2889 | -0.0013 | 3659.628480 | 13.6083 | 2715 | 13.6105 | 13.6106 | -0.0001 |
| | | 2440 | 13.2937 | 13.2950 | -0.0013 | | | 2720 | 13.6140 | 13.6137 | 0.0003 |
| | | 2445 | 13.2998 | 13.3010 | -0.0012 | **25** | 3.94351119 | 2725 | 13.6173 | 13.6168 | 0.0005 |
| | | 2450 | 13.3060 | 13.3070 | -0.0010 | | 4.35296986 | 2730 | 13.6205 | 13.6199 | 0.0006 |
| | | 2455 | 13.3122 | 13.3129 | -0.0007 | | | 2735 | 13.6234 | 13.6229 | 0.0005 |
| | | 2460 | 13.3184 | 13.3188 | -0.0004 | | | 2740 | 13.6261 | 13.6259 | 0.0002 |
| | | 2465 | 13.3247 | 13.3247 | 0.0000 | 2749.829382 | 13.6318 | 2745 | 13.6287 | 13.6289 | -0.0002 |
| | | 2470 | 13.3310 | 13.3306 | 0.0004 | 3621.198618 | 13.6318 | 2750 | 13.6312 | 13.6318 | -0.0006 |
| | | 2475 | 13.3372 | 13.3365 | 0.0007 | | | 2755 | 13.6336 | 13.6335 | 0.0001 |
| 2484.009390 | 13.3469 | 2480 | 13.3436 | 13.3423 | 0.0013 | **26** | 3.72963470 | 2760 | 13.6359 | 13.6350 | 0.0009 |
| | | | | | | | 4.09197173 | 2765 | 13.6381 | 13.6366 | 0.0015 |
| | | | | | | | | 2770 | 13.6402 | 13.6382 | 0.0020 |
| | | | | | | | | 2775 | 13.6422 | 13.6397 | 0.0025 |
| | | | | | | 2780.705225 | 13.6413 | 2780 | 13.6441 | 13.6411 | 0.0030 |
| | | | | | | 3590.322775 | 13.6413 | 2785 | 13.6460 | 13.6409 | 0.0051 |
| | | | | | | **27** | 3.40172802 | 2790 | 13.6478 | 13.6403 | 0.0075 |
| | | | | | | | 3.69140151 | 2795 | 13.6495 | 13.6397 | 0.0098 |
| | | | | | | 2800.782857 | 13.6390 | 2800 | 13.6512 | 13.6391 | 0.0121 |
| | | | | | | 3570.245143 | 13.6390 | 2805 | 13.6528 | 13.6332 | 0.0196 |
| | | | | | | **28** | 2.39586620 | | | | |
| | | | | | | | 2.46180014 | | | | |
| | | | | | | 2810.096720 | 13.6263 | 2810 | 13.6544 | 13.6264 | 0.0280 |
| | | | | | | 3560.931280 | 13.6263 | 2815 | 13.6559 | 13.6310 | 0.0249 |
| | | | | | | | | 2820 | 13.6574 | 13.6358 | 0.0216 |
| | | | | | | | | 2825 | 13.6588 | 13.6405 | 0.0183 |
| | | | | | | | | 2830 | 13.6601 | 13.6452 | 0.0149 |
| | | | | | | | | 2835 | 13.6613 | 13.6499 | 0.0114 |
| | | | | | | | | 2840 | 13.6625 | 13.6545 | 0.0080 |
| | | | | | | | | 2845 | 13.6636 | 13.6592 | 0.0044 |
| | | | | | | | | 2850 | 13.6646 | 13.6637 | 0.0009 |
| | | | | | | **29** | 4.30127189 | 2855 | 13.6655 | 13.6683 | -0.0028 |
| | | | | | | | 4.79178003 | 2860 | 13.6663 | 13.6728 | -0.0065 |
| | | | | | | | | 2865 | 13.6670 | 13.6772 | -0.0102 |
| | | | | | | | | 2870 | 13.6677 | 13.6817 | -0.0140 |
| | | | | | | | | 2875 | 13.6683 | 13.6861 | -0.0178 |
| | | | | | | | | 2880 | 13.6689 | 13.6904 | -0.0215 |
| | | | | | | | | 2885 | 13.6694 | 13.6948 | -0.0254 |
| | | | | | | | | 2890 | 13.6698 | 13.6991 | -0.0293 |
| | | | | Radius outer core = 3477,114454 | | 2893.913546 | 13.7024 | 2894 | 13.6700 | | |